\documentstyle[prd,preprint,tighten,aps,eqsecnum,floats,%
amssymb,amsfonts,newlfont,graphicx]{revtex}
\begin{document}
\draft
\preprint{
\begin{tabular}{r}
DFTT 32/99\\
hep-ph/9906275
\end{tabular}
}
\title{Neutrinoless double-$\beta$ decay with three or four neutrino mixing}
\author{Carlo Giunti}
\address{INFN, Sezione di Torino, and Dipartimento di Fisica Teorica,
Universit\`a di Torino,\\
Via P. Giuria 1, I--10125 Torino, Italy}
\maketitle
\begin{abstract}
Considering the scheme
with mixing of three neutrinos and a mass hierarchy
that can accommodate
the results of solar and atmospheric neutrino experiments,
it is shown that the results of solar neutrino experiments
imply a lower bound
for the effective Majorana mass in neutrinoless double-$\beta$ decay,
under the natural assumptions that massive neutrinos are Majorana particles
and there are no unlikely fine-tuned cancellations
among the contributions of the different neutrino masses.
Considering
the four-neutrino schemes that can accommodate also
the results of the LSND experiment,
it is shown that only one of them is compatible
with the results of neutrinoless double-$\beta$ decay experiments
and
with the measurement of the abundances of primordial elements
produced in Big-Bang Nucleosynthesis.
It is shown that in this scheme,
under the assumptions
that massive neutrinos are Majorana particles
and there are no cancellations
among the contributions of the different neutrino masses,
the results of the LSND experiment imply
a lower bound
for the effective Majorana mass in neutrinoless double-$\beta$ decay.
\end{abstract}


\section{Introduction}

Neutrino oscillations \cite{reviews,BGG-review-98,Kayser-review-99}
is one of the most intriguing phenomena of
today high-energy physics and one of the most promising
ways to explore the physics beyond the Standard Model.
At present
there are three experimental indications in favor of neutrino oscillations
that have been
obtained in solar neutrino experiments
(Homestake \cite{Homestake},
Kamiokande \cite{Kam-sun},
GALLEX \cite{GALLEX},
SAGE \cite{SAGE}
and
Super-Kamiokande \cite{SK-sun}),
in atmospheric neutrino experiments
(Kamiokande \cite{Kam-atm},
IMB \cite{IMB},
Super-Kamiokande \cite{SK-atm},
Soudan-2 \cite{Soudan2}
and
MACRO \cite{MACRO}))
and in the accelerator LSND experiment \cite{LSND,LSND-Moriond-99}.
On the other hand,
neutrinoless double-$\beta$ decay ($\beta\beta_{0\nu}$)
experiments \cite{double-beta,bb-exp,Baudis-99,genius}
and the experiments on the direct measurement of neutrino masses
\cite{tritium}
did not obtain any positive result so far
(see \cite{PDG98}).
The connection between the properties of neutrinos
that determine neutrino oscillations
(mass squared differences and neutrino mixing)
and neutrinoless double-$\beta$ decay
have been discussed in many papers
\cite{Petcov-Smirnov-94,BBGK,BGKP,BGKM-bb-98,BG-bb-98-99,BGGKP-bb-99,Vissani-bb,%
Minakata-Yasuda-bb,Fukuyama-bb,Adhikari-Rajasekaran-98,%
Branco-99,Barger-Whisnant-99-bb}.
In this paper we discuss some implications of
the latest results of neutrino oscillation experiments
for neutrinoless double-$\beta$ decay
and we show that under reasonable assumptions there
is a lower bound for the effective Majorana neutrino mass
measured in $\beta\beta_{0\nu}$ decay experiments.

If massive neutrinos are Majorana particles,
the matrix element
of $\beta\beta_{0\nu}$ decay is proportional to the effective Majorana mass
\begin{equation}
|\langle{m}\rangle|
=
\left|
\sum_{k}
U_{ek}^2
\,
m_{k}
\right|
\,,
\label{effective}
\end{equation}
where $U$ is the mixing matrix
that connects the flavor neutrino fields
$\nu_{\alpha L}$
($\alpha=e,\mu,\tau$)
to the fields $\nu_{kL}$ of neutrinos with masses $m_k$
through the relation
\begin{equation}
\nu_{\alpha L} = \sum_{k} U_{\alpha k} \, \nu_{kL}
\,.
\label{mixing}
\end{equation}
The present experimental upper limit for $|\langle{m}\rangle|$,
\begin{equation}
|\langle{m}\rangle|_{\mathrm{exp}}
\leq
0.2 - 0.4 \, \mathrm{eV}
\qquad
\mbox{(90\% CL)}
\,,
\label{exp-bb}
\end{equation}
has been obtained from the measurement of the half-life of $^{76}$Ge
in the Heidelberg--Moscow experiment
($ T_{1/2}^{0\nu}(^{76}\mathrm{Ge}) \geq 5.7 \times 10^{25} \, \mathrm{yr} $
at 90\% CL)
\cite{Baudis-99}.
The range of the upper bound
(\ref{exp-bb})
is due to the uncertainty of the theoretical calculation
of the nuclear matrix element
and has been obtained from the results of different
calculations using the
Quasiparticle Random Phase Approximation (QRPA)
\cite{QRPA,Simkovic-99}.
In particular,
the recent QRPA calculation in
\cite{Simkovic-99}
yields the rather stringent upper bound
$ |\langle{m}\rangle|_{\mathrm{exp}} \leq 0.27 \, \mathrm{eV} $.
On the other hand,
the Shell Model calculation in \cite{Caurier-96}
yields the loser bound
$ |\langle{m}\rangle|_{\mathrm{exp}} \leq 0.56 \, \mathrm{eV} $.
However,
the calculation of the
nuclear matrix element for the neutrinoless double-$\beta$ decay
of $^{76}$Ge
presented in \cite{Caurier-96}
has been truncated before reaching convergence
and the full calculation is expected to yield a more stringent upper bound
for $ |\langle{m}\rangle|_{\mathrm{exp}} $.
Therefore,
in the following we will consider the range in Eq. (\ref{exp-bb})
as a reliable estimate of the uncertainty of the experimental
upper bound for the effective Majorana mass
$ |\langle{m}\rangle| $.

The next generation of $\beta\beta_{0\nu}$ decay experiments
is expected to be sensitive to values of
$|\langle{m}\rangle|$
in the range $10^{-2} - 10^{-1}$ eV \cite{bb-exp}.
Values of
$|\langle{m}\rangle|$
as small as about
$10^{-3}$ eV may be reachable not far in the future \cite{genius}.

After the measurement in the Super-Kamiokande experiment
of an up-down asymmetry of $\mu$-like events
induced by atmospheric muon neutrinos,
the experimental evidence in favor of oscillations of atmospheric neutrinos
is widely considered to be beyond reasonable doubts
(see, for example, \cite{BGG-review-98,Lipari-99,Kayser-review-99}).
There are also convincing arguments in favor of a neutrino
oscillation explanation of the solar neutrino problem
(see, for example, \cite{BKS-sun-analysis-98,BGG-review-98,Berezinsky-99}).
Therefore,
in this paper we will consider first,
in Section \ref{Three neutrinos},
the implications for $\beta\beta_{0\nu}$ decay
in the scheme with mixing of three neutrinos and a mass hierarchy
that can accommodate
the results of atmospheric and solar neutrino experiments.
In Section \ref{Four neutrinos}
we consider the schemes with four massive neutrinos
that can accommodate all neutrino oscillation data,
including the LSND results in favor of
$\nu_\mu\to\nu_e$ and $\bar\nu_\mu\to\bar\nu_e$
oscillations that wait for independent confirmations
by other experiments \cite{KARMEN,LSND-check}.

\section{Three neutrinos with a mass hierarchy}
\label{Three neutrinos}

The results of solar and
atmospheric neutrino experiments indicate the existence of
a hierarchy of mass-squared differences
(see \cite{BKS-sun-analysis-98,SK-sun-analysis-99,Concha-sun-99,%
Concha-atm-98,FLMS-atm-98,SK-atm}):
\begin{equation}
\Delta{m}^2_{\mathrm{sun}}
\lesssim
10^{-4} \, \mathrm{eV}^2
\ll
10^{-3} \, \mathrm{eV}^2
\lesssim
\Delta{m}^2_{\mathrm{atm}}
\lesssim
10^{-2} \, \mathrm{eV}^2
\,,
\label{dm2-hierarchy}
\end{equation}
where
$\Delta{m}^2_{\mathrm{sun}}$
and
$\Delta{m}^2_{\mathrm{atm}}$
are the mass-squared differences
relevant for solar and atmospheric neutrino oscillations,
respectively.
A natural scheme that can accommodate this hierarchy
is the one with three neutrinos and a mass hierarchy,
\begin{equation}
\underbrace{
\overbrace{m_1 \ll m_2}^{\mathrm{sun}}
\ll m_3
}_{\mathrm{atm}}
\,.
\label{mass-hierarchy}
\end{equation}
In the framework of the hierarchical spectrum
(\ref{mass-hierarchy})
the mass-squared differences relevant
for the oscillations of solar and atmospheric neutrinos are
\begin{equation}
\Delta{m}^2_{\mathrm{sun}}
=
\Delta{m}^2_{21}
\equiv
m_2^2 - m_1^2
\simeq
m_2^2
\,,
\qquad
\Delta{m}^2_{\mathrm{atm}}
=
\Delta{m}^2_{31}
\equiv
m_3^2 - m_1^2
\simeq
m_3^2
\,.
\label{dm2-sun-atm}
\end{equation}
The mass hierarchy
(\ref{mass-hierarchy})
is predicted by the see-saw mechanism \cite{see-saw},
which
predicts also that the tree light massive neutrinos are
Majorana particles.
In this case
neutrinoless double-$\beta$ decay
is possible.

It has been shown in \cite{BGKM-bb-98,BG-bb-98-99,BGGKP-bb-99}
that the results of neutrino oscillation experiments
imply a rather stringent upper bound
(about $ 6 \times 10^{-3} \, \mathrm{eV}$)
for the effective Majorana
mass in neutrinoless double-$\beta$ decay
in the scheme with mixing of three neutrinos and a mass hierarchy.
In principle the effective Majorana mass (\ref{effective})
can be vanishingly small because of cancellations among the
contributions of the different mass eigenstates.
However,
since the neutrino masses and the elements of the neutrino mixing matrix
are independent quantities,
if there is a hierarchy of neutrino masses
such a cancellation would be the result of an unlikely fine-tuning,
unless some unknown symmetry is at work.
Here we consider the possibility that no such symmetry exist and
\emph{no unlikely fine-tuning operates
to suppress the effective Majorana mass} (\ref{effective}).
In this case we have
\begin{equation}
|\langle{m}\rangle|
\simeq
\max_k
|U_{ek}|^2
\,
m_{k}
\,.
\label{max}
\end{equation}
Let us define the absolute value of the contribution of the neutrino mass
$m_k$ to $|\langle{m}\rangle|$ as
\begin{equation}
|\langle{m}\rangle|_k
\equiv
|U_{ek}|^2
\,
m_{k}
\,.
\label{mk}
\end{equation}
In the following we will estimate the value of
$|\langle{m}\rangle|$
using the largest
$|\langle{m}\rangle|_k$
obtained from the results of neutrino oscillation experiments.

The results of the CHOOZ experiment \cite{CHOOZ}
and the Super-Kamiokande atmospheric neutrino data \cite{SK-atm}
imply that $|U_{e3}|^2$
is small
($|U_{e3}|^2 \lesssim 5 \times 10^{-2}$)
and there is an upper limit of about $6 \times 10^{-3}$ eV
for the contribution $|\langle{m}\rangle|_3$ to the effective
Majorana mass in $\beta\beta_{0\nu}$ decay
\cite{BGKM-bb-98,BG-bb-98-99,BGGKP-bb-99}.
Since there is no lower bound for $|U_{e3}|^2$
from experimental data,
$|\langle{m}\rangle|_3$
could be very small.

Hence,
the largest contribution to
$|\langle{m}\rangle|$
could come from
$
|\langle{m}\rangle|_2
\equiv
|U_{e2}|^2
\,
m_2
$.
Since in the framework of the scheme with
mixing of three neutrinos and a mass hierarchy
$\Delta{m}^2_{\mathrm{sun}} \simeq m_2^2$
and
$
|U_{e2}|^2
\simeq
\frac{1}{2}
\left(
1
-
\sqrt{ 1 - \sin^2 2\vartheta_{\mathrm{sun}} }
\right)
$
\cite{BG-98-dec},
where $\vartheta_{\mathrm{sun}}$
is the two-neutrino mixing angle used in the analysis of
solar neutrino data,
we have
\begin{equation}
|\langle{m}\rangle|_2
\simeq
\frac{1}{2}
\left(
1
-
\sqrt{ 1 - \sin^2 2\vartheta_{\mathrm{sun}} }
\right)
\sqrt{ \Delta{m}^2_{\mathrm{sun}} }
\,.
\label{m2}
\end{equation}

\begin{figure}
\begin{center}
\mbox{\includegraphics[bb=100 555 441 762,width=0.95\linewidth]{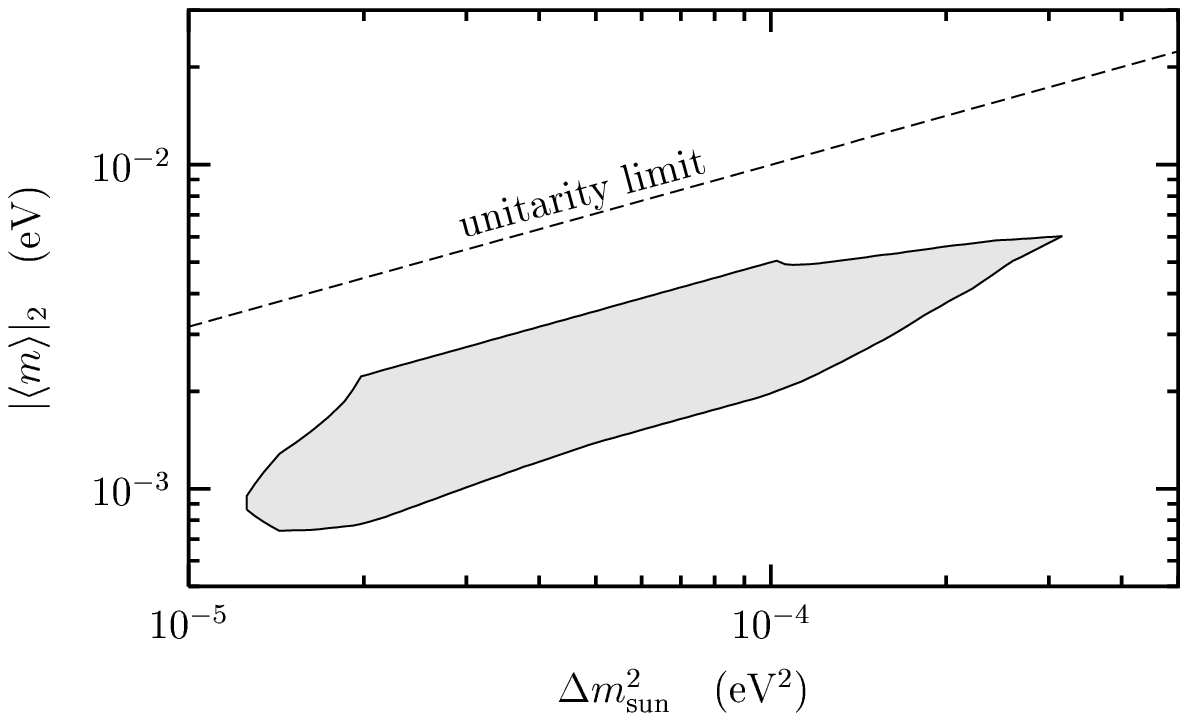}}
\end{center}
\caption{ \label{bb3}
The shadowed area shows the allowed range for
$|\langle{m}\rangle|_2$
as a function of
$\Delta{m}^2_{\mathrm{sun}}$
in the scheme with mixing of three neutrinos
with a mass hierarchy discussed in Section \ref{Three neutrinos},
in the case of the LMA-MSW solution of the solar neutrino problem.
The shadowed area has been
obtained using Eq. (\ref{m2})
and the allowed range for $\sin^2 2\vartheta_{\mathrm{sun}}$
given by the LMA-MSW region (99\% CL)
in the $\sin^2 2\vartheta_{\mathrm{sun}}$--$\Delta{m}^2_{\mathrm{sun}}$
plane presented
in Fig. 2 of Ref. \protect\cite{SK-sun-analysis-99}.
The dashed line
represents the unitarity limit
$ |\langle{m}\rangle|_2 \leq \sqrt{ \Delta{m}^2_{\mathrm{sun}} } $.}
\end{figure}

Solar neutrino data imply bounds for $\sin^2 2\vartheta_{\mathrm{sun}}$
and
$ \Delta{m}^2_{\mathrm{sun}} $.
In particular the large mixing angle MSW \cite{MSW} solution (LMA-MSW)
of the solar neutrino problem,
which seems to be favored by recent data \cite{BKS-99-LMA},
implies that \cite{SK-sun-analysis-99}
\begin{equation}
1.2 \times 10^{-5} \, \mathrm{eV}^2
\leq
\Delta{m}^2_{\mathrm{sun}}
\leq
3.1 \times 10^{-4} \, \mathrm{eV}^2
\,,
\qquad
0.58
\leq
\sin^2 2\vartheta_{\mathrm{sun}}
\leq
1.00
\label{LMA-MSW-12}
\end{equation}
at 99\% CL,
taking into account the total rates measured in solar neutrino experiments
and the day-night variations observed in the
Super-Kamiokande experiment \cite{SK-sun}.
Hence,
for the contribution of $m_2$ to the effective
Majorana mass we obtain
\begin{equation}
6 \times 10^{-4} \, \mathrm{eV}
\lesssim
|\langle{m}\rangle|_2
\lesssim
9 \times 10^{-3} \, \mathrm{eV}
\,.
\label{LMA-MSW-3}
\end{equation}
This estimate does not take into account the correlation between
$\Delta{m}^2_{\mathrm{sun}}$
and
$\sin^2 2\vartheta_{\mathrm{sun}}$.
The precise allowed range for
$|\langle{m}\rangle|_2$
as a function of
$\Delta{m}^2_{\mathrm{sun}}$
obtained with Eq. (\ref{m2}) from the LMA-MSW region (99\% CL)
in Fig. 2 of Ref. \cite{SK-sun-analysis-99}
is shown in Fig. \ref{bb3}.
The dashed line in Fig. \ref{bb3}
represents the unitarity limit
$ |\langle{m}\rangle|_2 \leq \sqrt{ \Delta{m}^2_{\mathrm{sun}} } $.
From Fig. \ref{bb3}
one can see that the LMA-MSW solution
of the solar neutrino problem implies that\footnote{
The upper limit
$
|\langle{m}\rangle|_2
\lesssim
3 \times 10^{-3} \, \mathrm{eV}
$
presented in \cite{BGGKP-bb-99}
has been obtained from the 90\% CL
LMA-MSW region in Fig. 8a of Ref. \cite{Concha-sun-99},
using Eq. (\ref{m2}).
The 99\% CL
LMA-MSW region in Fig. 8a of Ref. \cite{Concha-sun-99}
gives
$
|\langle{m}\rangle|_2
\lesssim
6 \times 10^{-3} \, \mathrm{eV}
$,
in agreement with the upper bound in Eq. (\ref{LMA-MSW-4}).
}
\begin{equation}
7.4 \times 10^{-4} \, \mathrm{eV}
\lesssim
|\langle{m}\rangle|_2
\lesssim
6.0 \times 10^{-3} \, \mathrm{eV}
\,.
\label{LMA-MSW-4}
\end{equation}

Assuming the absence of fine-tuned cancellations
among the contributions of the three neutrino masses
to the effective Majorana mass,
if $|U_{e3}|^2$ is very small and
$ |\langle{m}\rangle|_3 \ll |\langle{m}\rangle|_2$,
from Eqs. (\ref{max}) and (\ref{LMA-MSW-4}) we obtain
\begin{equation}
7 \times 10^{-4} \, \mathrm{eV}
\lesssim
|\langle{m}\rangle|
\lesssim
6 \times 10^{-3} \, \mathrm{eV}
\,.
\label{LMA-MSW-5}
\end{equation}
Hence, we see that,
\emph{assuming the absence of an
unlikely fine-tuned suppression of $|\langle{m}\rangle|$,
the results of solar neutrino experiments
give an indication of the value of the effective Majorana mass
in $\beta\beta_{0\nu}$ decay,
with a lower bound of about
$7 \times 10^{-4} \, \mathrm{eV}$
in the case of the LMA-MSW solution
of the solar neutrino problem}.
This bound is rather small,
but values of $|\langle{m}\rangle|$
of the order of
$10^{-3}$ eV,
indicated by the range (\ref{LMA-MSW-5}),
may be measurable in a not too far future \cite{genius}.

Also the small mixing angle MSW (SMA-MSW)
and the vacuum oscillation (VO)
solutions
of the solar neutrino problem imply allowed ranges for
$|\langle{m}\rangle|_2$,
but their values are much smaller than in the case of the LMA-MSW solution.
Using the 99\% CL allowed regions
obtained in \cite{BKS-sun-analysis-98}
from the analysis of the total rates measured in solar neutrino experiments
we have
\begin{eqnarray}
5 \times 10^{-7} \, \mathrm{eV}
\lesssim
|\langle{m}\rangle|_2
\lesssim
10^{-5} \, \mathrm{eV}
\null & \null \qquad \null & \null
\mbox{(SMA-MSW)}
\,,
\label{SMA-MSW}
\\
10^{-6} \, \mathrm{eV}
\lesssim
|\langle{m}\rangle|_2
\lesssim
2 \times 10^{-5} \, \mathrm{eV}
\null & \null \qquad \null & \null
\mbox{(VO)}
\,.
\label{VO}
\end{eqnarray}

\section{Four neutrinos}
\label{Four neutrinos}

If,
besides to the solar and atmospheric neutrino data,
also the results of the accelerator LSND experiment are taken into account,
at least three independent
mass-squared differences are needed.
This can be seen by considering
the general expression for the probability of
$\nu_\alpha\to\nu_\beta$
transitions in vacuum \cite{reviews,BGG-review-98,Kayser-review-99},
that can be written as
\begin{equation}
P_{\nu_\alpha\to\nu_\beta}
=
\left|
\sum_k
U_{{\alpha}k}^* \,
U_{{\beta}k} \,
\exp\left( - i \, \frac{ \Delta{m}^2_{kj} \, L }{ 2 \, E } \right)
\right|^2
\,,
\label{Posc}
\end{equation}
where
$ \Delta{m}^2_{kj} \equiv m_k^2-m_j^2 $,
$j$ is any of the mass-eigenstate indices,
$L$ is the distance between the neutrino source and detector
and $E$ is the neutrino energy.
The range of $L/E$ probed by each type of experiment is different:
$ L / E \gtrsim 10^{10} \, \mathrm{eV}^{-2} $
for solar neutrino experiments,
$ L / E \sim 10^{2} - 10^{3} \, \mathrm{eV}^{-2} $
for atmospheric neutrino experiments
and
$ L / E \sim 1 \, \mathrm{eV}^{-2} $
for the LSND experiment.
From Eq. (\ref{Posc}) it is clear that neutrino oscillations
occur in an experiment only if there is at least one
mass-squared difference
$\Delta{m}^2_{kj}$ such that
\begin{equation}
\frac{ \Delta{m}^2_{kj} \, L }{ 2 \, E }
\gtrsim 0.1
\label{cond1}
\end{equation}
(the precise lower bound depends on the sensitivity of the experiment)
in a significant part of the energy and source-detector distance
intervals of that experiment
(if the condition (\ref{cond1}) is not satisfied,
$
P_{\nu_\alpha\to\nu_\beta}
\simeq
\left|
\sum_k
U_{{\alpha}k}^* \,
U_{{\beta}k}
\right|^2
=
\delta_{\alpha\beta}
$).
Since the range of $L/E$ probed by the LSND experiment is the smaller one,
a mass-squared difference is needed for LSND oscillations:
\begin{equation}
\Delta{m}^2_{\mathrm{LSND}} \gtrsim 10^{-1} \, \mathrm{eV}^2
\,.
\label{dm2-LSND}
\end{equation}
Specifically,
the maximum likelihood analysis of the LSND data
in terms of two-neutrino oscillations
gives \cite{LSND-Moriond-99}
\begin{equation}
0.20 \, \mathrm{eV}^2
\leq
\Delta{m}^2_{\mathrm{LSND}}
\leq
2.0 \, \mathrm{eV}^2
\,.
\label{LSND-range}
\end{equation}

Furthermore,
from Eq. (\ref{Posc}) it is clear that 
a dependence of the oscillation probability
from the neutrino energy $E$
and the source-detector distance $L$
is observable only if there is at least one
mass-squared difference
$\Delta{m}^2_{kj}$ such that
\begin{equation}
\frac{ \Delta{m}^2_{kj} \, L }{ 2 \, E }
\sim 1
\,.
\label{cond2}
\end{equation}
Since
a variation of the transition probability as a function of neutrino energy
has been observed both in solar and atmospheric neutrino experiments
and the range of $L/E$ probed by each type of experiment is different,
two more mass-squared differences with different scales are needed:
\begin{eqnarray}
&&
\Delta{m}^2_{\mathrm{sun}} \sim 10^{-10} \, \mathrm{eV}^2
\qquad
\mbox{(VO)}
\,,
\label{dm2-sun}
\\
&&
\Delta{m}^2_{\mathrm{atm}} \sim 10^{-3} - 10^{-2} \, \mathrm{eV}^2
\,.
\label{dm2-atm}
\end{eqnarray}
The condition (\ref{dm2-sun}) for the solar mass-squared difference
$\Delta{m}^2_{\mathrm{sun}}$
has been obtained under the assumption of vacuum oscillations (VO). 
If the disappearance of solar $\nu_e$'s is due to the MSW effect \cite{MSW},
the condition
\begin{equation}
\Delta{m}^2_{\mathrm{sun}} \lesssim 10^{-4} \, \mathrm{eV}^2
\qquad
\mbox{(MSW)}
\label{dm2-sun-MSW}
\end{equation}
must be fulfilled
in order to have a resonance in the interior of the sun.
Hence,
in the MSW case
$\Delta{m}^2_{\mathrm{sun}}$
must be at least one order of magnitude smaller than
$\Delta{m}^2_{\mathrm{atm}}$.

The existence of three different scales
of neutrino mass-squared differences\footnote{
It is possible to ask if three different scales
of neutrino mass-squared differences are needed even
if the results of the Homestake solar neutrino experiment \cite{Homestake}
is neglected, allowing an energy-independent suppression of the solar $\nu_e$ flux.
The answer is that still the data cannot be fitted with only
two neutrino mass-squared differences
because an energy-independent suppression of the solar $\nu_e$ flux
requires large $\nu_e\to\nu_\mu$ or $\nu_e\to\nu_\tau$
transitions generated by $\Delta{m}^2_{\mathrm{atm}}$
or $\Delta{m}^2_{\mathrm{LSND}}$.
These transitions
are forbidden by the results of the Bugey \cite{Bugey} and CHOOZ \cite{CHOOZ}
$\bar\nu_e$ disappearance experiments
and by the non-observation of an up-down asymmetry
of $e$-like events in the Super-Kamiokande
atmospheric neutrino experiment \cite{SK-atm}.
I would like to thank S.T. Petcov for useful discussions about this point.}
imply that at least four light massive neutrinos must exist in nature.
Here we consider the schemes with four light and mixed neutrinos
\cite{four-models,four-phenomenology,%
Okada-Yasuda-97,BGG-AB,BGG-bounds+CP,BGGS-98-BBN,BGG-BBN-conf},
which constitute the minimal possibility
that allows to explain all the data of neutrino oscillation experiments.
In this case,
in the flavor basis the three active neutrinos $\nu_e$, $\nu_\mu$, $\nu_\tau$
are accompanied by a sterile neutrino $\nu_s$
that does not take part in
standard weak interactions. 

The existence of four light massive neutrinos
is a low-energy manifestation of physics
beyond the Standard Model (see, for example, \cite{sterile}).
In most theories beyond the Standard Model
neutrinos are naturally Majorana particles and
neutrinoless double-$\beta$ decay is allowed.
Therefore,
we see that the experimental evidences in favor of neutrino oscillations
indicate that neutrinos may be Majorana particles and
neutrinoless double-$\beta$ decay
is a concrete possibility.

It has been shown \cite{BGG-AB} that there are only
two schemes with four-neutrino mixing
that can accommodate the results of all neutrino oscillation
experiments:
\begin{equation}
\mbox{(A)}
\qquad
\underbrace{
\overbrace{m_1 < m_2}^{\mathrm{atm}}
<
\overbrace{m_3 < m_4}^{\mathrm{sun}}
}_{\mathrm{LSND}}
\,,
\qquad
\mbox{(B)}
\qquad
\underbrace{
\overbrace{m_1 < m_2}^{\mathrm{sun}}
<
\overbrace{m_3 < m_4}^{\mathrm{atm}}
}_{\mathrm{LSND}}
\,.
\label{AB}
\end{equation}
These two spectra are characterized by the presence of two pairs
of close masses separated by a gap of about 1 eV
which provides the mass-squared difference
$ \Delta{m}^2_{\mathrm{LSND}} = \Delta{m}^2_{41} $
responsible of the oscillations observed in the LSND experiment.
In the scheme A
$ \Delta{m}^2_{\mathrm{atm}} = \Delta{m}^2_{21} $
and
$ \Delta{m}^2_{\mathrm{sun}} = \Delta{m}^2_{43} $,
whereas in scheme B
$ \Delta{m}^2_{\mathrm{atm}} = \Delta{m}^2_{43} $
and
$ \Delta{m}^2_{\mathrm{sun}} = \Delta{m}^2_{21} $.

The results of
the short-baseline $\bar\nu_e$ disappearance experiment Bugey\cite{Bugey},
in which no indication in favor of neutrino oscillations was found,
imply that the mixing
of $\nu_e$ with the two
``heavy'' neutrinos $\nu_3$ and $\nu_4$
is large in scheme A
and
small in scheme B
\cite{BGG-AB,BGG-review-98}:
\begin{eqnarray}
1
-
\left(
|U_{e3}|^2
+
|U_{e4}|^2
\right)
&\lesssim&
3 \times 10^{-2}
\qquad
\mathrm{(A)}
\,,
\label{ce-small-A}
\\
|U_{e3}|^2
+
|U_{e4}|^2
&\lesssim&
3 \times 10^{-2}
\qquad
\mathrm{(B)}
\,,
\label{ce-small-B}
\end{eqnarray}
for
$\Delta{m}^2_{\mathrm{LSND}}$
in the LSND-allowed range (\ref{LSND-range}).
Therefore,
if scheme A is realized in nature
the effective Majorana mass in
$\beta\beta_{0\nu}$ decay
can be as large as
$ m_3 \simeq m_4 \simeq \sqrt{ \Delta{m}^2_{\mathrm{LSND}} }
\simeq 0.45 - 1.4 \, \mathrm{eV} $
\cite{BGKP,BGKM-bb-98,BG-bb-98-99,BGGKP-bb-99}.
On the other hand,
in scheme B
neutrinoless double-$\beta$ decay
is strongly suppressed
\cite{BGKM-bb-98,BG-bb-98-99,BGGKP-bb-99}.
In the following two subsections
we discuss some connections between the results of neutrino oscillation experiments
and
neutrinoless double-$\beta$ decay in the schemes A and B.

\subsection{Scheme A}
\label{Scheme A}

In the four-neutrino scheme A,
from Eq. (\ref{ce-small-A}) and the unitarity of the mixing matrix we have
$ |U_{e1}|^2 + |U_{e2}|^2 \lesssim 3 \times 10^{-2} $.
Therefore,
the contribution of the two light masses $m_1$ and $m_2$ to the
effective Majorana mass in $\beta\beta_{0\nu}$ decay can be neglected
and we have
\begin{equation}
|\langle{m}\rangle|
\simeq
\left|
U_{e3}^2 \, m_3
+
U_{e4}^2 \, m_4
\right|
\,,
\label{m-A}
\end{equation}
that implies the limits
\begin{equation}
\left|
|U_{e3}|^2 \, m_3
-
|U_{e4}|^2 \, m_4
\right|
\lesssim
|\langle{m}\rangle|
\lesssim
|U_{e3}|^2 \, m_3
+
|U_{e4}|^2 \, m_4
\,.
\label{m-A-limits-1}
\end{equation}
Neglecting the small difference between $m_3$ and $m_4$
and taking into account that
$ m_3 \simeq m_4 \simeq \sqrt{ \Delta{m}^2_{\mathrm{LSND}} } $,
we have
\begin{equation}
\left|
|U_{e3}|^2
-
|U_{e4}|^2
\right|
\sqrt{ \Delta{m}^2_{\mathrm{LSND}} }
\lesssim
|\langle{m}\rangle|
\lesssim
\left(
|U_{e3}|^2
+
|U_{e4}|^2
\right)
\sqrt{ \Delta{m}^2_{\mathrm{LSND}} }
\,.
\label{m-A-limits-2}
\end{equation}
Since the quantity
$
|U_{e3}|^2
+
|U_{e4}|^2
$
is large in scheme A (see Eq. (\ref{ce-small-A})),
we obtain
\begin{equation}
\left|
|U_{e3}|^2
-
|U_{e4}|^2
\right|
\sqrt{ \Delta{m}^2_{\mathrm{LSND}} }
\lesssim
|\langle{m}\rangle|
\lesssim
\sqrt{ \Delta{m}^2_{\mathrm{LSND}} }
\,.
\label{m-A-limits-3}
\end{equation}
Furthermore,
from the inequality (\ref{ce-small-A})
one can see that
the contribution of the mixing of $\nu_e$ with
$\nu_1$ and $\nu_2$
to the survival probability of solar electron neutrinos is negligible
and 
$|U_{e3}|^2$
and
$|U_{e4}|^2$
are related to the mixing angle $\vartheta_{\mathrm{sun}}$
obtained from the two-generation fit of solar neutrino experiments by
\begin{equation}
|U_{e3}|^2 \simeq \cos^2 \vartheta_{\mathrm{sun}}
\,,
\qquad
|U_{e4}|^2 \simeq \sin^2 \vartheta_{\mathrm{sun}}
\,.
\label{U-sun}
\end{equation}
Hence, the range (\ref{m-A-limits-3}) can be written as
\begin{equation}
\sqrt{ ( 1 - \sin^2 2\vartheta_{\mathrm{sun}} ) \, \Delta{m}^2_{\mathrm{LSND}} }
\lesssim
|\langle{m}\rangle|
\lesssim
\sqrt{ \Delta{m}^2_{\mathrm{LSND}} }
\,,
\label{m-A-limits-4}
\end{equation}
Let us emphasize that this allowed range for
$|\langle{m}\rangle|$
in scheme A
depends only on the assumption that massive neutrinos
are Majorana particles.

In the case of the SMA-MSW solution of the solar neutrino problem
(for both $\nu_e\to\nu_\tau$ or $\nu_e\to\nu_s$ transitions)
$\sin^2 2\vartheta_{\mathrm{sun}}$
is very small
($ \sin^2 2\vartheta_{\mathrm{sun}} \lesssim 10^{-2} $)
and we have
\begin{equation}
|\langle{m}\rangle|
\simeq
\sqrt{ \Delta{m}^2_{\mathrm{LSND}} }
\simeq 0.45 - 1.4 \, \mathrm{eV}
\qquad
(\mbox{SMA-MSW})
\,.
\label{m-A-SMA}
\end{equation}
Hence,
the experimental upper bound (\ref{exp-bb})
indicates that
\emph{the SMA-MSW solution of the solar neutrino problem
is disfavored in scheme A}.

Furthermore,
the upper bound $ N_\nu^{\mathrm{BBN}} < 4 $
for the effective number of neutrinos
in Big-Bang Nucleosynthesis (BBN)
(see, for example, \cite{Schramm-Turner-98})
implies that
\cite{Okada-Yasuda-97,BGGS-98-BBN,BGG-BBN-conf}
\begin{equation}
|U_{s1}|^2 + |U_{s2}|^2 \lesssim 10^{-2}
\,,
\label{cs-small-A}
\end{equation}
in scheme A.
The analysis of recent astrophysical
data yields the upper bound\footnote{
The bound $N_\nu^{\mathrm{BBN}} \leq 3.2$ \cite{Burles-99}
implies that
$ |U_{s1}|^2 + |U_{s2}|^2 \lesssim 5 \times 10^{-4} $
\cite{BGGS-98-BBN,BGG-BBN-conf}
in scheme A!
}
$N_\nu^{\mathrm{BBN}} \leq 3.2$
at 95\% CL
\cite{Burles-99},
although the issue is still rather controversial
(see \cite{Olive-99,Sarkar-99}).
The inequalities (\ref{ce-small-A}) and (\ref{cs-small-A}),
together with the unitarity of the mixing matrix,
imply that
the oscillations of solar neutrinos
occur mainly in the
$ \nu_e \to \nu_s $
channel
\cite{BGGS-98-BBN,BGG-BBN-conf}.
In this case,
the analysis of solar neutrino data
in terms of two-generation $ \nu_e \to \nu_s $
oscillations is valid in the four-neutrino scheme A
if the usual two-generation mixing parameters
$\Delta{m}^2$ and $\vartheta$
are identified, respectively,
with
$\Delta{m}^2_{\mathrm{sun}}=\Delta{m}^2_{43}$ and $\vartheta_{\mathrm{sun}}$
defined in Eq. (\ref{U-sun})
(from Eqs. (\ref{ce-small-A}), (\ref{U-sun}), (\ref{cs-small-A})
and the unitarity of the mixing matrix we obtain
$|U_{s3}|^2 \simeq \sin^2\vartheta_{\mathrm{sun}}$
and
$|U_{s4}|^2 \simeq \cos^2\vartheta_{\mathrm{sun}}$).
The results of the analyses of solar neutrino data
in terms of two-generation $ \nu_e \to \nu_s $
oscillations
show that
only the SMA-MSW solution is allowed
\cite{BKS-sun-analysis-98,SK-sun-analysis-99,Concha-sun-99}.
Therefore,
comparing Eqs. (\ref{exp-bb}) and (\ref{m-A-SMA})
we conclude that
\emph{the scheme A
is disfavored by the experimental upper bound
$ |\langle{m}\rangle| $
and the BBN bound $ N_\nu^{\mathrm{BBN}} < 4 $}.

Summarizing,
the data from oscillation experiments,
from neutrinoless double-$\beta$ decay experiments
and
from the measurement of the abundances of primordial elements
indicate that,
\emph{among all the possible four-neutrino schemes
there is only one allowed, scheme B}.

\subsection{Scheme B}
\label{Scheme B}

In scheme B,
the BBN upper bound $ N_\nu^{\mathrm{BBN}} < 4 $
implies that
\cite{BGGS-98-BBN,BGG-BBN-conf}
\begin{equation}
|U_{s3}|^2 + |U_{s4}|^2 \lesssim 10^{-4}
\,.
\label{cs-small-B}
\end{equation}
From this inequality, Eq. (\ref{ce-small-B})
and the unitarity of the mixing matrix
it follows that
the oscillations of solar neutrinos
occur mainly in the
$ \nu_e \to \nu_s $
channel.
Therefore,
the analysis of solar neutrino data
in terms of two-generation $ \nu_e \to \nu_s $
oscillations is valid in the four-neutrino scheme B
if the usual two-generation mixing parameters
$\Delta{m}^2$ and $\vartheta$
are identified,
respectively,
with
$\Delta{m}^2_{\mathrm{sun}}=\Delta{m}^2_{21}$ and $\vartheta_{\mathrm{sun}}$
defined by
\begin{equation}
|U_{e1}|^2 \simeq \cos^2 \vartheta_{\mathrm{sun}}
\,,
\qquad
|U_{e2}|^2 \simeq \sin^2 \vartheta_{\mathrm{sun}}
\label{U-sun-B}
\end{equation}
(from Eqs. (\ref{ce-small-B}), (\ref{cs-small-B}), (\ref{U-sun-B})
and the unitarity of the mixing matrix we obtain
$|U_{s1}|^2 \simeq \sin^2\vartheta_{\mathrm{sun}}$
and
$|U_{s2}|^2 \simeq \cos^2\vartheta_{\mathrm{sun}}$).
Since
the results of the analyses of solar neutrino data
in terms of two-generation $ \nu_e \to \nu_s $
oscillations
\cite{BKS-sun-analysis-98,SK-sun-analysis-99,Concha-sun-99}
show that
only the SMA-MSW solution is allowed,
with
$ 10^{-3} \lesssim \sin^2 2\vartheta_{\mathrm{sun}} \lesssim 10^{-2} $,
we have
$
2.5 \times 10^{-4}
\lesssim
|U_{e2}|^2
\lesssim
2.5 \times 10^{-3}
$.
Therefore,
in scheme B we have
\begin{equation}
|U_{e1}|^2 \simeq 1
\,,
\qquad
|U_{e2}|^2
\,,
|U_{e3}|^2
\,,
|U_{e4}|^2
\ll 1
\,.
\label{Ue-B}
\end{equation}

In the scheme B there are two possibilities:
a quasi-degenerate mass spectrum
\begin{equation}
\underbrace{
\overbrace{m_1 < m_2}^{\mathrm{sun}}
\lesssim
\overbrace{m_3 < m_4}^{\mathrm{atm}}
}_{\mathrm{LSND}}
\qquad
\mbox{(BD)}
\label{BD}
\end{equation}
or a mass hierarchy
\begin{equation}
\underbrace{
\overbrace{m_1 < m_2}^{\mathrm{sun}}
\ll
\overbrace{m_3 < m_4}^{\mathrm{atm}}
}_{\mathrm{LSND}}
\qquad
\mbox{(BH)}
\label{BH}
\end{equation}

If the quasi-degenerate mass spectrum BD is realized in nature,
it is clear that
from Eqs. (\ref{effective}) and (\ref{Ue-B})
we have
\begin{equation}
|\langle{m}\rangle| \simeq m_1
\,.
\label{m-BD}
\end{equation}
In this case,
the experimental upper bound (\ref{exp-bb})
implies that
\begin{equation}
m_1
\lesssim
0.2 - 0.4 \, \mathrm{eV}
\,.
\label{m1-BD}
\end{equation}
The observation of
neutrinoless double-$\beta$ decay
by the next generation of experiments,
which are sensitive to values of
$|\langle{m}\rangle|$
in the range $10^{-2} - 10^{-1}$ eV \cite{bb-exp},
together with the confirmation of the four-neutrino scheme B
by neutrino oscillation experiments,
will provide an evidence in favor of the
quasi-degenerate scheme BD.

If the hierarchical mass spectrum BH is realized in nature,
the absence of unlikely fine-tuned cancellations
between the contributions of
$m_1$, $m_2$
and
$m_3$, $m_4$
to the effective Majorana mass (\ref{effective})
implies that
\begin{equation}
|\langle{m}\rangle|
\simeq
\max\!\left[ |\langle{m}\rangle|_{12} , |\langle{m}\rangle|_{34} \right]
\,,
\label{max-4}
\end{equation}
with
\begin{eqnarray}
|\langle{m}\rangle|_{12}
&\equiv&
\left| U_{e1}^2 \, m_1 + U_{e2}^2 \, m_2 \right|
\,,
\label{m12-def}
\\
|\langle{m}\rangle|_{34}
&\equiv&
\left| U_{e3}^2 \, m_3 + U_{e4}^2 \, m_4 \right|
\,.
\label{m34-def}
\end{eqnarray}

From Eq. (\ref{Ue-B}),
if
$m_1 \simeq m_2$
we have
$ |\langle{m}\rangle|_{12} \simeq m_2 $
and the contribution of
$|\langle{m}\rangle|_{12}$
to
$|\langle{m}\rangle|$
could be sizable.
On the other hand,
if
$m_1 \ll m_2$
we have
$ |\langle{m}\rangle|_{12} \ll m_2 $
and
the contribution of
$|\langle{m}\rangle|_{12}$
to
$|\langle{m}\rangle|$
is strongly suppressed.
In this case,
if there are no unlikely fine-tuned cancellations
between the contributions of $m_1$ and $m_2$ to
$|\langle{m}\rangle|_{12}$,
we have
$|\langle{m}\rangle|_{12} \simeq |\langle{m}\rangle|_{2}$
with $|\langle{m}\rangle|_{2}$
in the range (\ref{SMA-MSW}).
In any case,
at present it is not known if
$m_1 \simeq m_2$
or
$m_1 \ll m_2$
and we cannot infer the contribution of
$|\langle{m}\rangle|_{12}$
to
$|\langle{m}\rangle|$.

Let us consider now the contribution
$|\langle{m}\rangle|_{34}$
to the effective Majorana mass.
In principle it is possible that
$|\langle{m}\rangle|_{34}=0$
if
$ U_{e3}^2 \, m_3 + U_{e4}^2 \, m_4 = 0 $,
\textit{i.e.}
\begin{equation}
|U_{e3}|^2 \, m_3 = |U_{e4}|^2 \, m_4
\label{bb-cond1}
\end{equation}
and
\begin{equation}
\left|
\arg(U_{e3}) - \arg(U_{e4})
\right|
=
\pi/2
\label{bb-cond2}
\end{equation}
(the condition (\ref{bb-cond2}) is satisfied if CP is conserved
and $\nu_3$ and $\nu_4$ have opposite CP parities
\cite{CP-parity,BGKP,BGG-review-98,BGGKP-bb-99}).
However,
even if $m_3 \simeq m_4$,
since
$|U_{e3}|^2 + |U_{e4}|^2 \ll 1$
(see Eq. (\ref{ce-small-B})),
there is no reason to have
$|U_{e3}|^2 \simeq |U_{e4}|^2$.
On the other hand,
the explanation of the atmospheric neutrino data
with $\nu_\mu\to\nu_\tau$ oscillations
\cite{Kam-atm,IMB,SK-atm,Soudan2,MACRO},
that is favored by the latest Super-Kamiokande data
\cite{Learned-JHU99},
requires a large mixing in the
$\nu_\mu,\nu_\tau$--$\nu_3,\nu_4$
sector
which could be related to the fact that
$|U_{\mu3}|^2 + |U_{\mu4}|^2$
and
$|U_{\tau3}|^2 + |U_{\tau4}|^2$
are large (close to one)
and
$m_3 \simeq m_4$.

Therefore,
in the following we will assume that
$|U_{e3}|^2$ and $|U_{e4}|^2$
have different orders of magnitude.
In this case,
the contribution of $m_3$ and $m_4$
to the effective Majorana mass
is given by
\begin{equation}
|\langle{m}\rangle|_{34}
\simeq
d_e \, \sqrt{ \Delta{m}^2_{\mathrm{LSND}} }
\,,
\label{m34}
\end{equation}
where we have taken into account that
$ m_3 \simeq m_4 \simeq \sqrt{ \Delta{m}^2_{41} }
= \sqrt{ \Delta{m}^2_{\mathrm{LSND}} } $
and we have defined
\begin{equation}
d_\alpha
\equiv
\sum_{k=3,4}
|U_{\alpha k}|^2
\qquad
(\alpha=e,\mu,\tau,s)
\,.
\label{calphaB}
\end{equation}

It has been shown in \cite{BGG-AB}
that $d_e$ is small in scheme B:
\begin{equation}
d_e
\leq
a_e^{\mathrm{Bugey}}
\,,
\label{de}
\end{equation}
with
\begin{equation}
a_e^{\mathrm{Bugey}}
=
\frac{1}{2} \left( 1 - \sqrt{ 1 - \sin^2 2\vartheta_{\mathrm{Bugey}} } \right)
\,.
\label{aeBugey}
\end{equation}
Here $\sin^2 2\vartheta_{\mathrm{Bugey}}$
is the upper value of the two-neutrino mixing parameter
$\sin^2 2\vartheta$
obtained from the Bugey exclusion curve \cite{Bugey}
as a function of
$\Delta{m}^2 = \Delta{m}^2_{\mathrm{LSND}}$,
where $\Delta{m}^2$ is the two-neutrino mass-squared difference
used in the analysis of the Bugey data
(the upper bound (\ref{ce-small-B})
has been obtained from the inequality (\ref{de})
restricting $\Delta{m}^2_{\mathrm{LSND}}$
in the LSND-allowed range (\ref{LSND-range})).

\begin{figure}
\begin{center}
\mbox{\includegraphics[bb=100 555 447 761,width=0.95\linewidth]{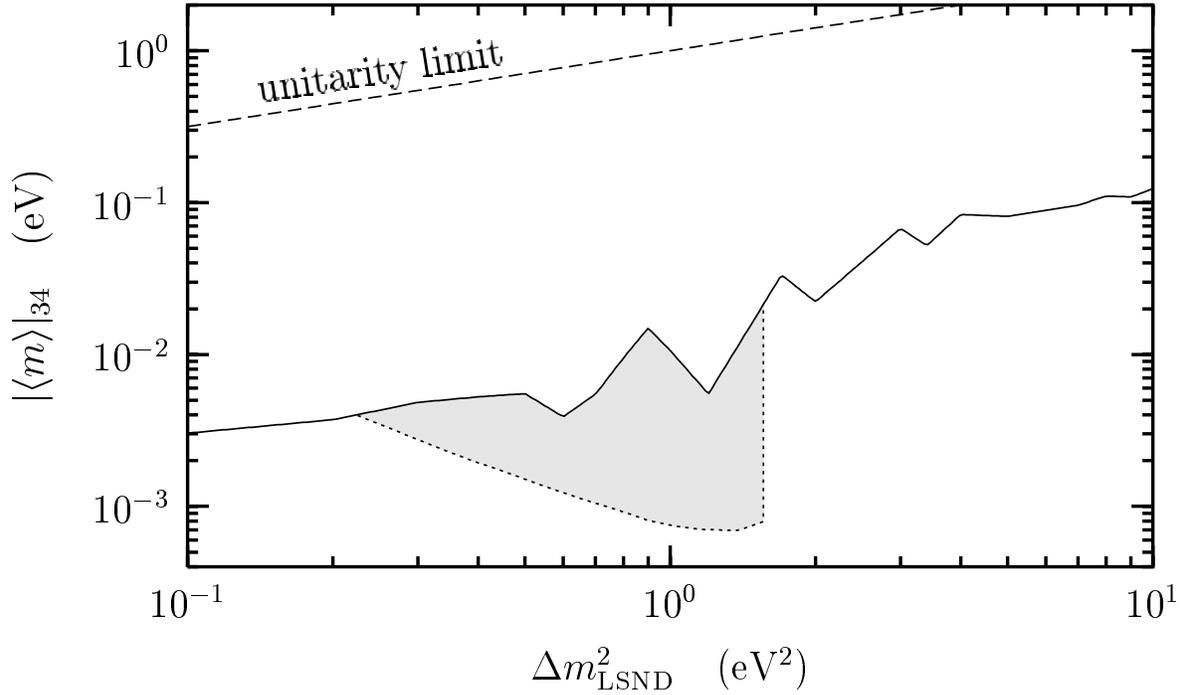}}
\end{center}
\caption{ \label{bb4}
The shadowed region shows the allowed range for
$|\langle{m}\rangle|_{34}$
as a function of
$\Delta{m}^2_{\mathrm{LSND}}$
in the four-neutrino scheme BH (see Eq. (\ref{BH})).
The solid line
represents the upper bound in Eq. (\ref{B4-bound})
and the dotted line
represents the lower bound in Eq. (\ref{m34-lb}).
The dashed line
represents the unitarity limit
$ |\langle{m}\rangle|_{34}
\leq \sqrt{ \Delta{m}^2_{\mathrm{LSND}} }
$.}
\end{figure}

From Eqs. (\ref{m34}) and (\ref{de}),
for
$|\langle{m}\rangle|_{34}$
we obtain the upper bound \cite{BGKM-bb-98,BG-bb-98-99,BGGKP-bb-99}
\begin{equation}
|\langle{m}\rangle|_{34}
\lesssim
a_e^{\mathrm{Bugey}} \, \sqrt{ \Delta{m}^2_{\mathrm{LSND}} }
\,,
\label{B4-bound}
\end{equation}
The numerical value of this upper bound
as a function of
$\Delta{m}^2_{\mathrm{LSND}}$
is depicted by the solid line in Fig. \ref{bb4}.
The dashed line in Fig. \ref{bb4}
represents the unitarity limit
$ |\langle{m}\rangle|_{34}
\leq \sqrt{ \Delta{m}^2_{\mathrm{LSND}} }
$.

The amplitude
$
A_{\mu e}
=
4
\left|
\sum_{k=3,4}
U_{e k} \, U_{\mu k}^*
\right|^2
$
of short-baseline $\bar\nu_\mu\to\bar\nu_e$
oscillations in scheme B is bounded by \cite{BGG-AB}
\begin{equation}
A_{\mu e} \leq 4 \, d_e \, d_\mu
\,.
\label{amue-ub}
\end{equation}
Since $d_\mu$ is large in scheme B \cite{BGG-AB}, we have
\begin{equation}
d_e
\geq
\frac{ A_{\mu e}^{\mathrm{min}} }{ 4 }
\,,
\label{ceBmin}
\end{equation}
where
$A_{\mu e}^{\mathrm{min}} $
is the minimum value of
$A_{\mu e}$
measured in the LSND experiment.
The physical reason of this lower bound for $d_e$
is that $\nu_e$ must have some mixing with $\nu_3$ and/or $\nu_4$
in order to generate the oscillations observed in the LSND
experiment.

Taking into account Eq. (\ref{m34}),
the inequality (\ref{ceBmin}) leads to the lower bound
\begin{equation}
|\langle{m}\rangle|_{34}
\gtrsim
\frac{ A_{\mu e}^{\mathrm{min}} }{ 4 }
\ \sqrt{ \Delta{m}^2_{\mathrm{LSND}} }
\,.
\label{m34-lb}
\end{equation}
The numerical value of this lower bound
as a function of
$\Delta{m}^2_{\mathrm{LSND}}$
is shown in Fig. \ref{bb4}
by the dotted curve that,
together with the solid line obtained from the upper bound
(\ref{B4-bound})
defines an allowed region
in the
$\Delta{m}^2_{\mathrm{LSND}}$--$|\langle{m}\rangle|_{34}$ plane
(shadowed area).
From Fig. \ref{bb4}
one can see that
\begin{equation}
6.9 \times 10^{-4} \, \mathrm{eV}
\lesssim
|\langle{m}\rangle|_{34}
\lesssim
2.1 \times 10^{-2} \, \mathrm{eV}
\,.
\label{m34-bounds}
\end{equation}

Summarizing,
in the framework of the scheme BH in Eq. (\ref{BH})
we have made three assumptions:
(i)
massive neutrinos are Majorana particles,
(ii)
there is no
unlikely fine-tuned cancellations
between the contributions of
$m_1$, $m_2$
and
$m_3$, $m_4$
to the effective Majorana mass $|\langle{m}\rangle|$,
(iii)
the two small elements $U_{e3}$ and $U_{e4}$
of the neutrino mixing matrix have different orders of magnitude.
Under these reasonable assumptions,
we have obtained the following allowed range for
the effective Majorana mass in $\beta\beta_{0\nu}$ decay:
\begin{equation}
7 \times 10^{-4} \, \mathrm{eV}
\lesssim
|\langle{m}\rangle|
\lesssim
2 \times 10^{-2} \, \mathrm{eV}
\,,
\label{m-bounds}
\end{equation}
from the contribution of
$m_3$ and $m_4$
alone
(the effective Majorana mass could be even larger than
$ 2 \times 10^{-2} \, \mathrm{eV} $
if
$ m_1 \simeq m_2 \gtrsim 2 \times 10^{-2} \, \mathrm{eV} $).
Such values of the effective Majorana mass
could be measured by future
$\beta\beta_{0\nu}$ decay experiments \cite{bb-exp,genius}.

\section{Conclusions}

We have derived lower limits
for the the effective Majorana mass in neutrinoless double-$\beta$ decay
in the scheme with mixing of three neutrinos and a mass hierarchy
[Eq. (\ref{mass-hierarchy})]
under the natural assumptions that
massive neutrinos are Majorana particles
and there are no large cancellations among the
contributions of the different neutrino masses.
If there is a hierarchy of neutrino masses,
large cancellations are unlikely
(unless an unknown symmetry is at work),
because they require a fine-tuning among the
values of the neutrino masses and the elements of the
neutrino mixing matrix,
which are independent quantities.

Under the only assumption that massive neutrinos
are Majorana particles,
we have shown that,
among all the possible four-neutrino schemes
that can accommodate the results of solar and atmospheric experiments
and the results of the LSND experiment,
only the scheme B [Eq. (\ref{AB})]
is compatible
with the experimental results on neutrinoless double-$\beta$ decay
and
the measurements of the cosmic abundances of elements
produced in Big-Bang Nucleosynthesis.
In the scheme B there are two possibilities:
the quasi-degenerate mass spectrum BD [Eq. (\ref{BD})]
and
the hierarchical mass spectrum BH [Eq. (\ref{BH})].

If the quasi-degenerate four-neutrino scheme BD
is realized in nature,
neutrinoless double-$\beta$ decay should be observed by the
next generation of experiments,
which will be sensitive to
$|\langle{m}\rangle| \sim 10^{-2} - 10^{-1}$ eV.

In the framework of the hierarchical four-neutrino scheme BH,
we have shown that there is a lower bound for the
effective Majorana mass in $\beta\beta_{0\nu}$ decay,
under the assumptions that
massive neutrinos are Majorana particles,
there are no large cancellations among the
contributions of $m_1,m_2$ and $m_3,m_4$
and
the two small elements
$U_{e3}$ and $U_{e4}$
of the neutrino mixing matrix
have different orders of magnitude.

We hope that the indications presented here in favor of a lower bound,
albeit small,
for the effective Majorana mass in neutrinoless double-$\beta$ decay
will encourage the development of future
$\beta\beta_{0\nu}$
experiments.

\newpage

\newpage

\end{document}